# Scale Ratio Tuning of Group Based Job Scheduling in HPC Systems

D. S. Lyakhovets,[1, *]    A. V. Baranov,[1, **]    and P. N. Telegin[1, ***]

[1]Joint Supercomputer Center of the Russian Academy of Sciences – Branch of Federal State Institution "Scientific Research Institute for System Analysis of the Russian Academy of Sciences", Leninsky prospect, 32a, Moscow, 119334, Russia

Abstract—During the initialization of a supercomputer job, no useful calculations are performed. A high proportion of initialization time results in idle computing resources and less computational efficiency. Certain methods and algorithms combining jobs into groups are used to optimize scheduling of jobs with high initialization proportion. The article considers the influence of the scale ratio setting in algorithm for the job groups formation, on the performance metrics of the workload manager. The study was carried out on the developed by authors Aleabased workload manager model. The model makes it possible to conduct a large number of experiments in reasonable time without losing the accuracy of the simulation. We performed a series of experiments involving various characteristics of the workload. The article represents the results of a study of the scale ratio influence on efficiency metrics for different initialization time proportions and input workflows with varying intensity and homogeneity. The presented results allow the workload managers administrators to set a scale ratio that provides an appropriate balance with contradictory efficiency metrics.

## 1. INTRODUCTION

The research of the scale ratio influence on the performance metrics of the workload manager was performed by authors using the workload manager model. The model is based on the Alea simulator, and implements the Packet algorithm [1].

Supercomputer tasks are typically organized as jobs. A job is an entity of work created by user. It contains parallel program(s), and requires input data and certain computer resources to enable its execution.

A variety of jobs submitted by different users form job queues. These queues are processed by special software like SLURM, PBS, LSF. This kind of software is called workload manager or batch system or job management system (JMS). In the Joint Supercomputer Center of the Russian Academy of Sciences) [2] the SUPPZ JMS is used for over 20 years.

Stages of job execution include initialization, job run, job completion. All stages except job run are overheads. The overhead time reduction increases efficiency. The high proportion of overhead results from either long initialization/completion time or short job runtime. Low overhead proportion usually have no negative impact on JMS efficiency metrics.

In this article we investigate overheads caused by job initialization, which includes all preparatory actions before the job run, including computing resources allocation. In this article we review

---
*
**

E-mail: abaranov@jscc.ru

\*\*\* E-mail: ptelegin@jscc.ru

moldable jobs with linear speed-up [3] and constant initialization time. For such jobs, the initialization time depends on the job type only and does not depend on the amount of allocated resources. The execution time is in inverse proportion to the amount of allocated computing resources, i.e., more resources used, less job execution is.

To reduce overhead in our previous work [4] we suggested a group based job scheduling method for JMS and discussed its implementation for the SUPPZ. We showed that job grouping considerably increases efficiency for the jobs with high initialization proportion. In our work [1] we described implementation of group based job scheduling method in Alea simulator as Packet algorithm. This significantly reduced the duration of experiments. Besides that, there were defined the initialization thresholds when grouping can result in measurable benefit on the efficiency metrics for homogeneous and inhomogeneous workloads with underload, load and overload modes. These modes correspond to the calculated load rates 0.85, 0.9, and 0.95.

An important setting for the Packet algorithm is the scale ratio. The impact of the scale ratio on efficiency metrics can be inconsistent. The chosen scale ratio can reduce the job time in queue, which is good for the users, but it can reduce the useful utilization, which is bad for system administrators, and vice versa. In general, the choice of scale ratio can improve some efficiency metrics, but decrease the other ones. The JMS administrator needs recommendations for setting the scale ratio to find a balance between conflicting efficiency metrics, taking into account the interests of the users and the efficient use of computing resources.

The remaining question was the influence of scale ratio on JMS efficiency. To answer the question we conducted a series of experiments to determine the influence of the scale ratio on efficiency metrics for different workflows. Previously developed model based on Alea made it possible to conduct lots of experiments in reasonable time.

This article is further development of the works [1], [4], and [5]. We present our research results of the scale ratio effect on job grouping.

At the moment, there is no method for configuring the scale ratio for the existing JMS with implemented group based job scheduling. We investigate, whether the characteristics of the workflow influence the choice of the scale ratio.

Let us consider current papers on jobs grouping and JMS modeling.

## 2. RELATED WORK

In the article [4] we reviewed different classes of jobs with high proportion of initialization. This kind jobs are frequent in photo and video processing, motion analysis, lighting rendering after adding or removing objects, genome sequencing, drug testing, nuclear physics, and bioinformatics. Huge overhead may be inherent in jobs that involve preparation and launch of a virtual platform, including the use of container virtualization [6].

Let us consider jobs with the identical initialization as jobs of the same type. Jobs of the same type from one user can be combined in one group (meta-job). For a meta-job, initialization can be performed once, whereafter the jobs from the group are run one by one, thus resulting in overhead reduction.

Paper [7] presents jobs classes with runtime less than 1 second (rapid tasks), 5 seconds (fast tasks), 30 seconds (medium tasks), 60 seconds (long tasks). The proposed approach combines all jobs that should be executed on a single physical core into one scheduler task, packing all individual tasks into a loop. This reduces scheduling overhead by avoiding re-dispatching and initializing multiple jobs. Various job types are not supported.

In the work [8] authors note the dynamic provision of virtual machines using Aneka, a platform for developing scalable applications based upon Azure cloud provider (Microsoft) takes 20% of the average overhead for initialization.

In the previous work [9] it was shown that the average overhead for launching the job in SUPPZ was 10 seconds. Hence, for a job with 1 second execution time the initialization proportion will make 91%, for 5 seconds execution time – 67%, for 30 seconds – 25%, for 60 seconds – 14%.

Most authors consider job grouping only for a single jobs, and their approaches are not applicable to a workload with different job types [10]–[14].

Optimizing the job queue is related to the job-shop scheduling problem. However its use in scheduling supercomputer jobs is hampered by strict restrictions on the number of scheduled jobs, computing nodes, or on the offline operation of the algorithm, when all the jobs are available for scheduling and no new jobs are added. The work [15] presents static job scheduling for a meal manufacturing process in collective catering with up to 110 jobs.

The work [16] presents the review of results of about 50 papers related to scheduling with setup times, basically it is static scheduling. Only seven papers consider approaches to continuously incoming jobs. Notably, the dynamic schedule is built up with a number of other limitations like no more than 21 computing nodes, which makes those methods inapplicable for modern highperformance systems.

The work [4] proposes group based job scheduling that allows to solve online job scheduling problem and to group jobs by types, without limits for the number of nodes and input jobs. This work does not determine the initialization threshold proportion when the grouping algorithm outperforms the widely used in JMS backfill algorithm.

One of frequently applied methods for the case is simulation modeling [17]. Researchers can get access to various model options like MONARC [18], Alea [19], OptorSim [20], WorkflowSim [21], Batsim [22], and SLURM simulator [23]. We have chosen the Alea [19] simulator. Using Alea, various scheduling algorithms are implemented like a common queue (FCFS), backfill algorithm and others, and it is possible to implement a custom job scheduling algorithm.

The work [24] presents the results of experiments proving that SUPPZ simulator is more accurate than Alea model. But Alea allows to handle long experiments (weeks of real workload) shortly (dozens of minutes).

In our study, the job type is a part of the job itself, so it is known beforehand. This separate research area is discussed in a variety of works. For example, when a computing system does not provide a job type, it is yet possible to use clustering to determine job types. In the paper [25] proposed an accurate smart task clustering method that uses both static information on the executable files and dynamic data about the behavior of applications during their execution.

Besides ordinary rigid jobs, there are widely used malleable jobs which dynamically reconfigure resources during runtime. In [26] there is presented a batch-system simulator supporting the combined scheduling of rigid and malleable jobs.

New simulators are created, particularly the work [27] describes a new simulator to approximate the actual Slurm scheduling realization.

According to the review the scale ratio influence on JMS efficiency metrics for different workflows is still an open question. To answer this question, we built a simulation model based on Alea, and ran simulations for various workloads.

### 3. JMS EFFICIENCY METRICS

There is no common set of efficiency metrics for analyzing JMS. The goal of the study determines which efficiency metrics should be used. We measured following efficiency metrics [28]:
1. Full utilization of computing resources.
2. Useful utilization of computing resources. Computing nodes are considered idle during resource initialization when calculating the useful utilization.
3. Job waiting time in queue (queue time).
4. Queue length.

For all metrics, an average value for a period is calculated. We used the period from the experiment start to the last job submit. In all experiments the last job was submitted about 4 days after the experiment start. After last submit the experiment continued for some time longer until all jobs were executed, but this period was not used for efficiency metrics calculation.

We also calculated median value for a job queue time in addition to average one.

System administrator needs information on full and useful utilization that reflect the efficiency computing resources use. Queue length is a simple way for administrator to detect how loaded a computing system is at a given time. It is useful for the user to know the job average queue time. Using this metric, the user can estimate how quickly his job can start.

## 4. PREVIOUS SCALE RATIO STUDIES

In previous works [5] and [28] the main goal was to determine whether a group based job scheduling has a positive impact on JMS efficiency metrics. To do this, it was necessary to estimate approximately the impact of job grouping settings driven by scale ratio on the efficiency metrics of the JMS. To achieve the goal, we performed experiments to study the influence of the scale ratio on efficiency metrics with the use of SUPPZ simulator with virtual computing nodes. This simulator performs experiments in real time, therefore a week workload requires a week of computational experiment. This limits the number of experiments, which makes it possible to study the changeable parameters only in a narrow range of values. So, only scale ratio values in the range 0.25-2.0 were studied for one input stream, with 5 different values of the initialization proportion. Scale ratio significantly affects the average job queue time (Fig. 1).

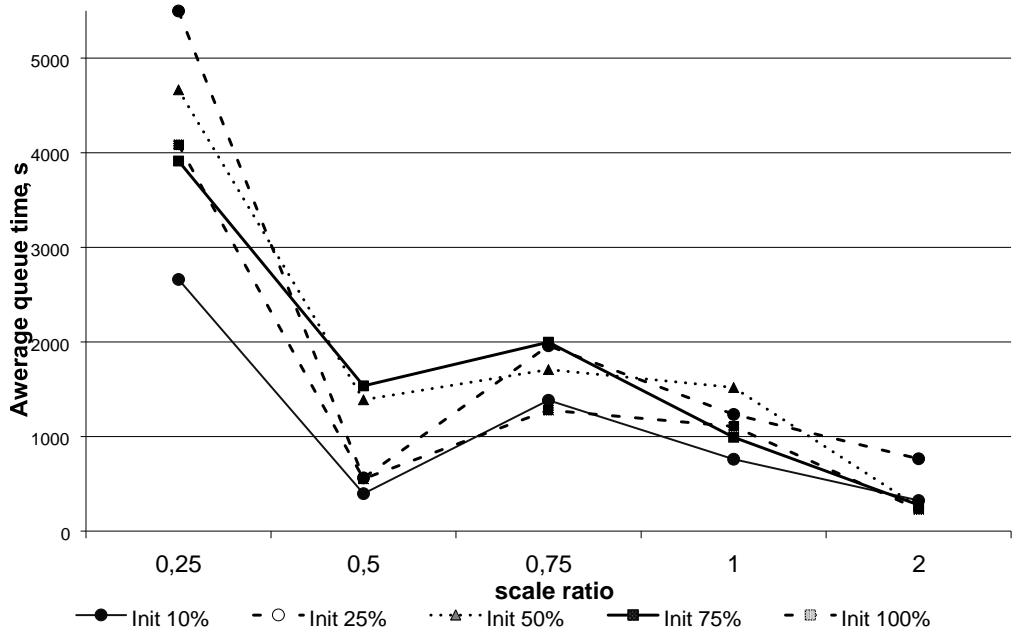

Figure 1. Influence of scale ratio on average job queue time for the initialization proportion from 10 to 100% and 8 job types. Aggregate data based on [28]

The effect of the scale ratio on the useful utilization is less significant (Fig. 2). For example, for the 50% initialization proportion, the useful utilization varies from 81% (scale ratio 0.25) to 87% (scale ratio 2). There is no stable dependence of the scale ratio on useful utilization.

Approximate estimates above show that the scale ratio influences efficiency metrics, but it is impossible to make recommendations for setting the scale ratio. The real time experiment speed of the used simulator resulted the study in a relatively small range of values.

In further research, we analyzed the existing models and developed our own model based on Alea [1]. This significantly reduced the time of experiments.

In [24] we compared accuracy in various JMS models. The model based on the Alea simulator provides accuracy comparable to the SUPPZ model with virtual nodes, while significantly reducing the time of the experiment. As a result, the Alea-based model is a promising tool for conducting extensive experiments.

In [5] and [28] we changed the average job processing time in the generated workload to obtain a more homogeneous input workflow. This method is suitable for testing the impact of grouping on efficiency metrics, but it may not fully reflect the impact on the actual workload. In our previous

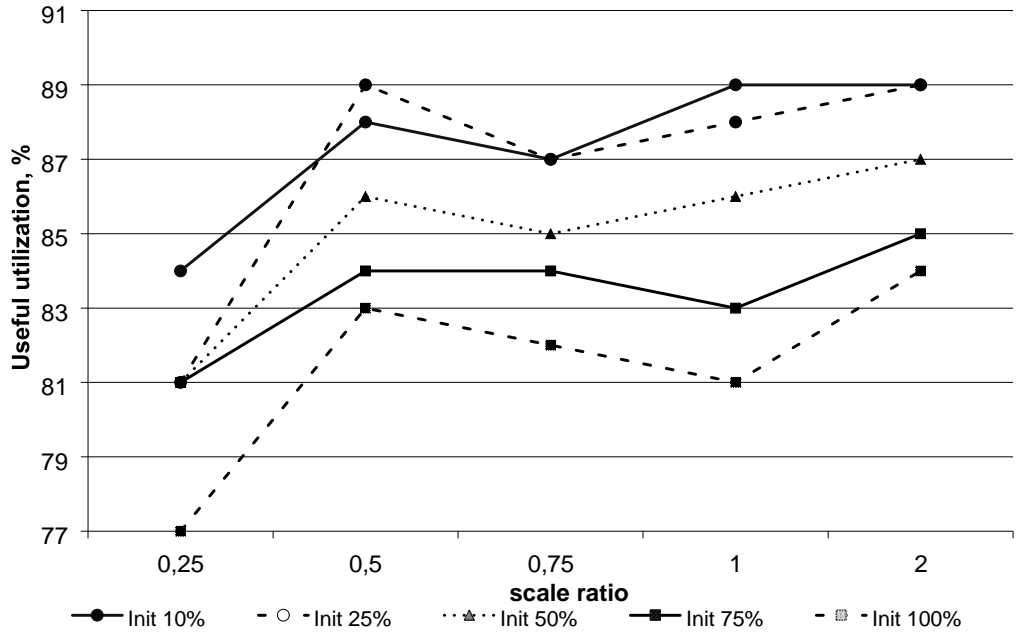

Figure 2. Influence of scale ratio on the average job queue time for initialization proportion from 10 to 100% and 8 job types. Aggregate data based on [28]

work [1] we used the generator from [29] and defined a starting threshold where grouping can result in measurable benefit on the efficiency metrics for homogeneous and heterogeneous workloads in underload, load and overload modes. These modes correspond to the workload with 0.85, 0.9 and 0.95 calculated load. We continue to study these 6 flows in this article.

The question remained is impact of scale ratio on efficiency metrics of JMS. To address this, our goal is to do a series of experiments with workloads described in [1]. We determine the impact of the scale ratio on JMS efficiency metrics for each workflow. Using the Alea model, it is possible to conduct a large series of experiments in reasonable time.

## 5. SCALE RATIO IN PACKET ALGORITHM

Scale ratio $k$ is a dimensionless coefficient representing scaling jobs onto computing resources. Job group will be executed on a number of nodes so that job group execution time $k$ times exceeds its initialization time. When possible, a job group is started on a number of computing nodes so that the total processing time of all jobs in the group is $k$ times greater than the initialization time. The higher the scale ratio is the less resources the job group runs on to reduce initialization overhead.

The Packet algorithm manages a separate queue for each job type, forms jobs of same type into groups and determines the required resources amount for each group. Each type is associated with some software environment initialization that is some initialization program and subsequent execution of jobs of this type.

Step 1. When the computing node is released, calculate the weights of $h$ queues.

Step 2. Select the queue with the largest weight from all non-empty queues $W(T_i) = \max(W(T_j))$, $j \in (1,h)$. $W(T_j)$ is calculated from the following formula $W(T_j) = C_j P_j (1 + T_j^{cur}/T_j^{max})$, where $W(T_j)$ is weight of the queue $T_j$; $C_j$ is advisability of grouping queue $j$, ratio of sum of all jobs execution times in the queue $j$ to $j$ initialization time, where $j \in (1,h)$, $C_j = \sum_{i=1}^{n} e_i/S_j$; $e_i$ is execution time of job $i$ on 1 node, $i \in (1,n)$; $s_j$ is initialization time of type $j$; $P_j$ is job type priority; $T_j^{cur}$ is queue time of the very first job in the queue $T_j$; $T_j^{max}$ is max queue time in queue $T_j$.

Step 3. Form a job group of all $n$ jobs from the selected queue.

Step 4. Determine the number of $m_{group}$ nodes for processing a job group, with appropriate scale ratio.

Let us consider the number of nodes for job group execution. Value $m_{threshold}$ for $n$ jobs of type $j$ is the following $m_{threshold} = \sum_{i=1}^{n} e_i/(ks_j)$, $k$ is specified scale ratio;

Job group is executed on $m_{group} = \min(m_{threshold}, m_{free})$ nodes. If there are not enough free nodes then the job group will be executed on all free nodes. In this case job group executions time exceeds initialization time by more than $k$ times. This results in additional reduction in overhead costs.

Step 5. Submit the generated package to the JMS for execution on $m_{group}$ nodes.

Let's consider an example. Let the initialization time for a certain queue be 1 time unit (1 minute for example). The group includes jobs with a total duration of 4 minutes when run on 1 computing node. With a scale ratio of 0.5, the package should run on such a number of computing nodes that the processing time is 1/2 of the initialization time, that is, 0.5. This will require 8 nodes, and the job group will take 1 minute to initialize and 0.5 minutes to complete (Fig. 3). With scale ration of 1, the group run on 4 nodes, and the processing time will be 1 minute. When $k = 2$, then 2 computing nodes will be selected, and the processing time will be 2 minutes. For $k = 4$ we have 1 node selected and the processing time will be 4 minutes.

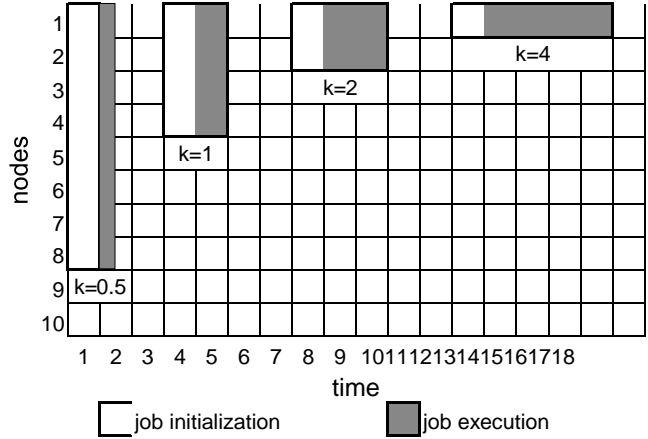

Figure 3. Impact of scale ratio on the number of computing nodes for processing a job group

The higher the scale ratio, the less computing resources the job group runs on. This reduces initialization overhead and increases useful utilization. Running a group on a small number of computing nodes can result in decrease of the utilization when there are not enough jobs to load all the computing nodes. In the case of intensive submit of the same type jobs, a case may happen when all incoming jobs will be formed into a single group and run on 1 computing module. The remaining nodes will be idle.

The lower the scale ratio, the more computing resources the job group runs on. This increases the initialization overhead. In this case, a larger number of computing nodes is used, which can have a positive effect on the utilization. In the case with be a significant number of the same type jobs in the queue they will be formed into one job group and run on the most of the computing nodes to ensure a given scale ratio. Jobs of other types will wait in the queue for a long time, since there are not enough nodes to run them.

Scale ratio significantly affects the efficiency metrics [28]. At the same time, it is impossible to estimate beforehand this influence, since it can differ significantly depending on the intensity and homogeniosity of the workflow. The problem of experimental study of the scale ratio influence on efficiency metrics for various workloads becomes urgent.

## 6. JOB MANAGEMENT SYSTEM MODEL FOR STUDY SCALE RATIO

In the article [4] we used the SUPPZ model with virtual supercomputing nodes.

In [5], we described implementation of the Packet algorithm in Alea. We improved the model and used it to study the impact of scale ratio. The new JMS model is based on Alea simulator, and implements the Packet algorithm. This is shown in Fig. 4. The input of the model is an investigated workflow. The changeable parameters of the model are nodes count, workflow and scale ratio. The result of Alea modeling is the output (or evaluated workflow) where Alea defined the start and end time for each job. Efficiency metrics analyzer processes the output and calculates the efficiency of scheduling.

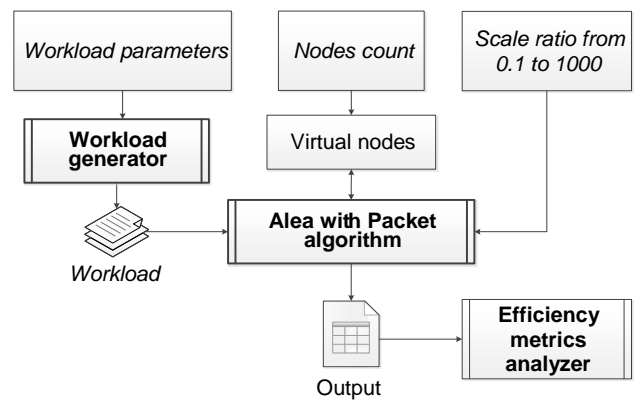

Figure 4. JMS model for scale ratio study

The workflow generator allows to explore an arbitrary JMS. In HPC research, there are often performed studies on different workflows [30]. It is possible to use both a real job workflow from the logfile, and a generated one. When generating a workflow, it is possible to use both public statistics of different JMS and your own statistical information.

We used 6 workflows, described in the work [1], which were created basing on the workload generator from [29]. This generator creates a workload statistically similar to a real workload of a supercomputer. The generated jobs parameters are submit time, runtime, required number of supercomputer nodes. Each of the parameters is modeled by a random variable from some distribution. The type and the distribution's parameters are based on the statistical analysis of several supercomputer logs for long period.

Each workflow consisted of 5000 jobs coming over 4 days. The first group of workflows was generated using the original generator from [29]. The second group was more homogeneous due to the use of a modified generator. In each group, there were three workloads generated, each of them differed in the calculated load of computing resources. A workload with load 0.85 illustrates an underloaded JMS mode.

For the loaded mode, a workload with 0.9 load was used. To study the JMS behavior in overload mode a workload with 0.95 load was chosen.

Nodes count depends on the computer installation. We used the values of 500 for heterogeneous workflows and 100 for homogeneous ones.

We made a series of experiments for 6 workflows with different values of the scale ratio. For each workflow, there were carried 37 experiments for the scale ratio ranging from 0.1 to 1.0 with a step of 0.1, from 1 to 10 with a step of 1, from 10 to 100 with a step of 10, and from 100 to 1000 with a step of 100. It was experimentally determined that the scale ratio value over 50 does not influence the efficiency metrics for any of the workflows. At the same time, for both homogeneous and heterogeneous workflows, we detected the equal impact of the scale ratio on efficiency metrics. So, we illustrate this effect based on more uniform workflows. Let us call the least loaded flow as Workload0.85, the medium-loaded flow as Workload0.90 and the most intense flow will be called Workload0.95.

The initialization time was set as a constant value for all jobs for each experiment. We calculate an average initialization time proportion S as follows

$$S = \frac{\sum_{i=1}^{n} s_i}{\sum_{i=1}^{n} s_i + \sum_{i=1}^{n} e_i},$$

where $s_i$ is $i^{th}$ job initialization time, $e_i$ is $i^{th}$ job runtime, $n = 5000$ is number of jobs.

We conducted experiments for each workflow with different initialization time proportions varying from 5% to 50%. Thus, the total number of experiments was "the number of workflows" multiplied by the "number of experiments with varying scale ratio" and multiplied by the "number of experiments with varying initialization proportion", a total of 6 × 37 × 6 = 1332 experiments.

By creating the Alea-based model, we got an opportunity to study a much more workflows and large ranges of various JMS parameters. We applied this model to study the influence of the scale ratio on the JMS efficiency metrics.

## 7. EXPERIMENTAL RESULTS

Let us consider the Workload0.85 with an average initialization time of 5%. The average queue time graph looks like presented in Fig. 5. One can see that as the scale ratio increases, the average queue time for a job in the queue decreases. To construct the graph we used the results of 23 experiments (one for each value of the scale ratio) for parameters above and various scale ratio.

Low values of the scale ratio significantly influence all efficiency metrics. When the scale ratio is above 20 the average job queue time reaches a plateau and further scale ratio growth does not influence it.

The median queue time of the Workload0.85 with an average initialization time of 5% gives the same graph form, but the plateau is reached much earlier. For a scale ratio of only 8, the median queue time is 0.

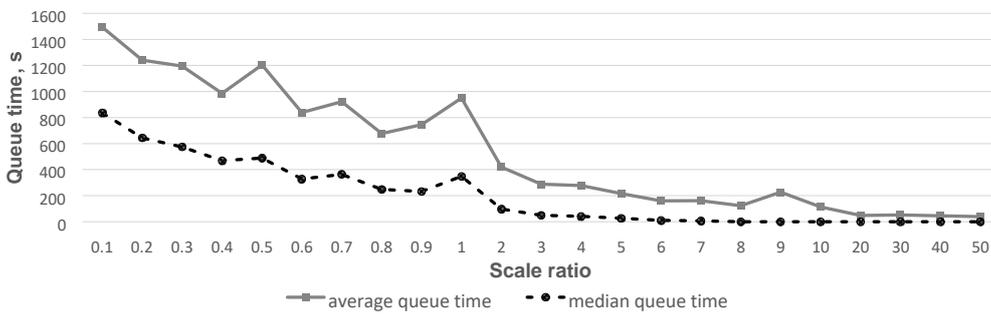

Figure 5. Average and median queue time for Workload0.85 with 5% initialization proportion. The less the better

The graph of the average queue length in all experiments performed looks similar to the average queue time, so we will not consider this indicator further. Let us consider an example for 5% of the average of initialization time proportion (Fig. 6). Plateau is reached when the scale ratio is 20.

When the average initialization time proportion is 50%, the median queue time decreases quicker and becomes equal to zero already when the scale ratio is 4 (Table 1, Fig. 7).

Consider the influence of the scale ratio on the average queue time for different initialization proportions (Fig. 8). Experiments with initialization proportions 5% shown in dotted line with squares, 10% – dotted line with circles, 30% – dashed line with squares, 50% – solid line with squares. As the scale ratio increases, the average queue time decreases with any average initialization proportion. When the value of the scale ratio is above 20, average queue time goes to plateau. Experiments were carried out for initialization proportions of 5%, and from 10% to 50% with 10% steps. For easier demonstration, the values of 20% and 40% are not shown on the graph.

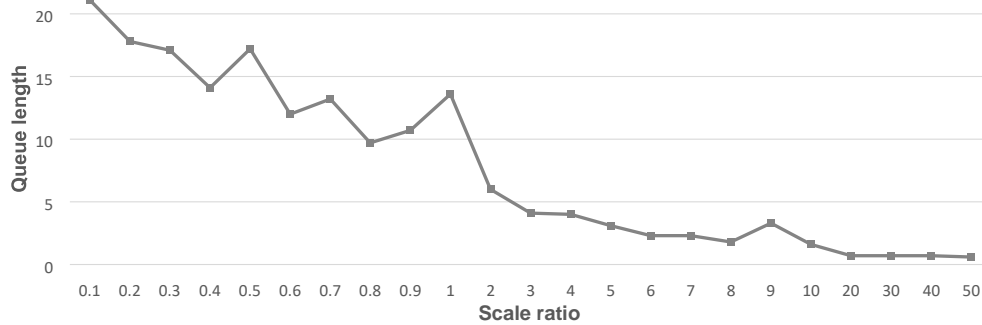

Figure 6. Average queue length for Workload0.85 with 5% initialization proportion. The less the better

Table 1. Average and median queue time for Workload0.85 for initialization proportion of 50% and scale ratio from 0.1 to 0.5

| Scale ratio | 0.1 | 0.2 | 0.3 | 0.4 | 0.5 |
|---|---|---|---|---|---|
| Average queue time, s | 9448 | 4505 | 8361 | 453 | 436 |
| Median queue time, s | 8187 | 3671 | 7306 | 203 | 189 |

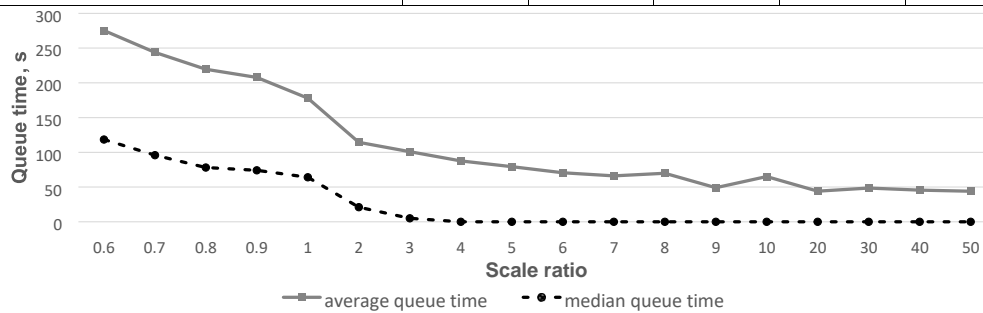

Figure 7. Average and median queue time for Workload0.85 with 50% initialization proportion. The less the better

When the scale ratio is n the range from 0.1 to 0.5, the average queue time is too long to be presented in the same graph, so we will present this fragment in a Table 2.

One can see in the graph that the lines for a 5% initialization proportion limit the graph at the top, and for the 50% proportion they limit the graph at the bottom. This trend was observed for all the considered workflows.

Let us consider the modified workflow of an average intensity Workload0.90 (Table 3, Fig. 9). The average queue time is growing, but the trend in the graph does not change: as the ratio increases, the average job queue time decreases, and at a certain ratio value it stabilizes.

Table 2. Average queue time for Workload0.85 for initialization proportion from 5% to 50% with scale ratio from 0.1 to 0.5

| Initialization proportion | Scale Ratio | | | | |
|---|---|---|---|---|---|
| | 0.1 | 0.2 | 0.3 | 0.4 | 0.5 |
| 5% | 1493 | 1241 | 1195 | 986 | 1204 |
| 10% | 1577 | 1488 | 1129 | 1055 | 915 |
| 30% | 7019 | 864 | 759 | 508 | 416 |
| 50% | 9448 | 4505 | 8361 | 453 | 436 |

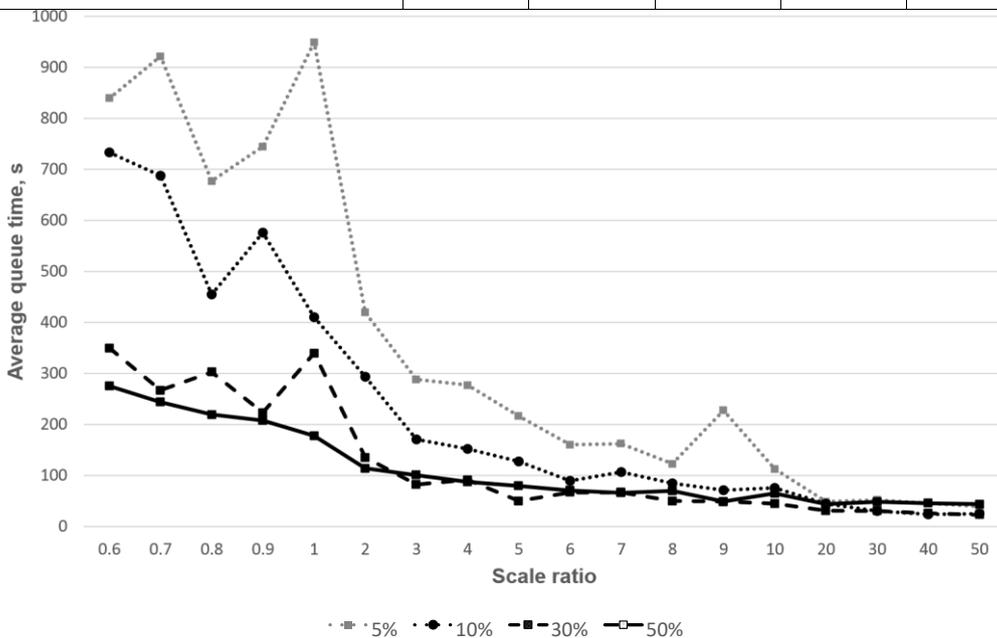

Figure 8. Average queue time for Workload0.85 for initialization proportion from 5% to 50%. The less the better

Table 3. Average queue time for Workload0.90 for initialization proportion of 5% and 50%, scale ratio ranges from 0.1 to 0.5

| Initialization proportion | Scale Ratio | | | | |
|---|---|---|---|---|---|
| | 0.1 | 0.2 | 0.3 | 0.4 | 0.5 |
| 5% | 2722 | 2423 | 1611 | 1921 | 1272 |

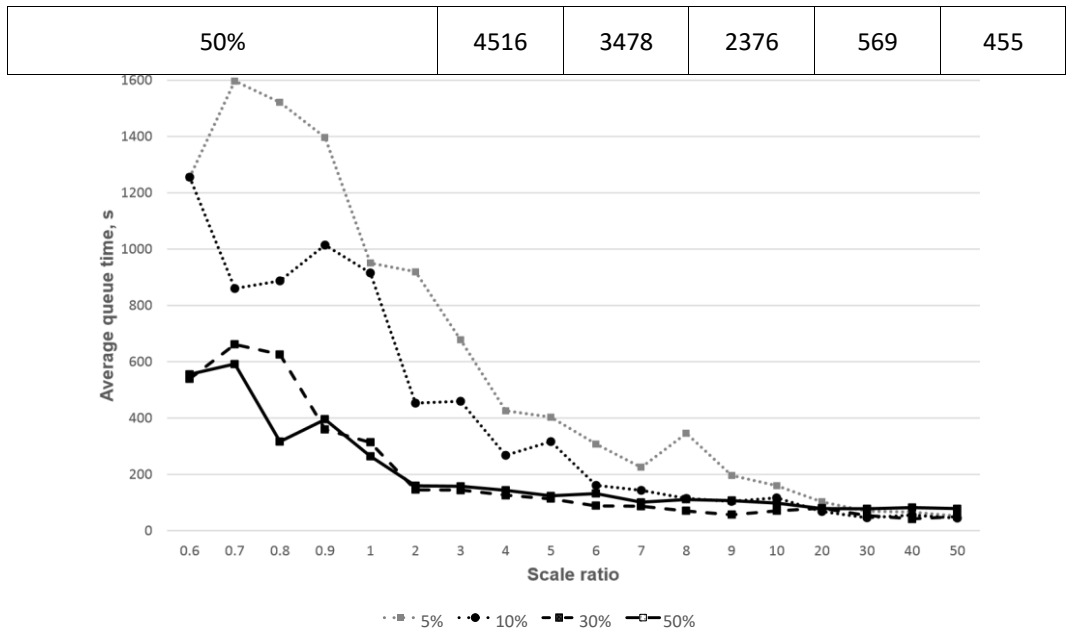

| 50% | 4516 | 3478 | 2376 | 569 | 455 |

Figure 9. Average queue time for Workload0.90 initialization proportion of 5%, 10%, 30%, and 50%. The less the better

Moreover, for each of the workflow and any initialization proportion, as the scale ratio increases, the average and median queue time decreases. After a certain value of the ratio, the queue time stabilizes at plateau and does not change further.

Let us fix the average initialization proportion. Let us consider the influence of the workflow intensity on the average queue time. Fig. 10 presents a comparative graph of the average queue time for all three workloads with different intensity for an average initialization proportion of 5%.

You can see a similar trend with an increase in the scale ratio towards a decrease in the queue time and further stabilization. When the scale ratio reaches of 50, the graph goes into a plateau. As the utilization increases, the average queue time also increases. At the same time, increasing the scale ratio leads to a decrease in queue time.

The scale ratio significantly influences the queue time. With an increase in the scale ratio for any of the explored workflows, the values of the average and median queue time stabilize at a certain value. The numerical value of the scale ratio, the average and median queue time depend on the workflow characteristics. For different input streams, the influence of the scale ratio efficiency metrics demonstrates a similar dependence. Presumably, a similar trend will be observed for any long workflow while keeping the distribution and its parameters for each of the job characteristics. As part of our study, we significantly varied the workflows basing on calculated load and uniformity, and the trend lasted.

Let's consider the influence of the scale ratio on the utilization of computing resources. For Workload0.85, for any initialization proportion, full utilization decreases with increasing scale ratio (Fig. 11).

Let us fix the average initialization proportion. Let us consider the influence of the workflow on full utilization. In Fig. 12 a comparative graph is presented for full utilization for three workloads of different intensities for a 5% average initialization proportion. For low values of the scale ratio, high values of the full utilization of computing resources are observed. As the scale ratio increases, the full utilization of computing resources decreases.

As the scale ratio increases, the full utilization decreases, this negatively effects the efficiency metrics of the JMS. One can see a conflict between efficiency metrics. To reduce the average queue time, the scale ratio should be reduced, but this has a negative impact on the full utilization of computing resources.

Let us consider the influence of the scale ratio on the useful utilization of computing resources. For Workload0.85 with an average initialization time of 5%, the useful utilization changes slightly (Fig. 13). When the scale ratio varies the useful utilization for a fixed initialization proportion changes within the statistical error.

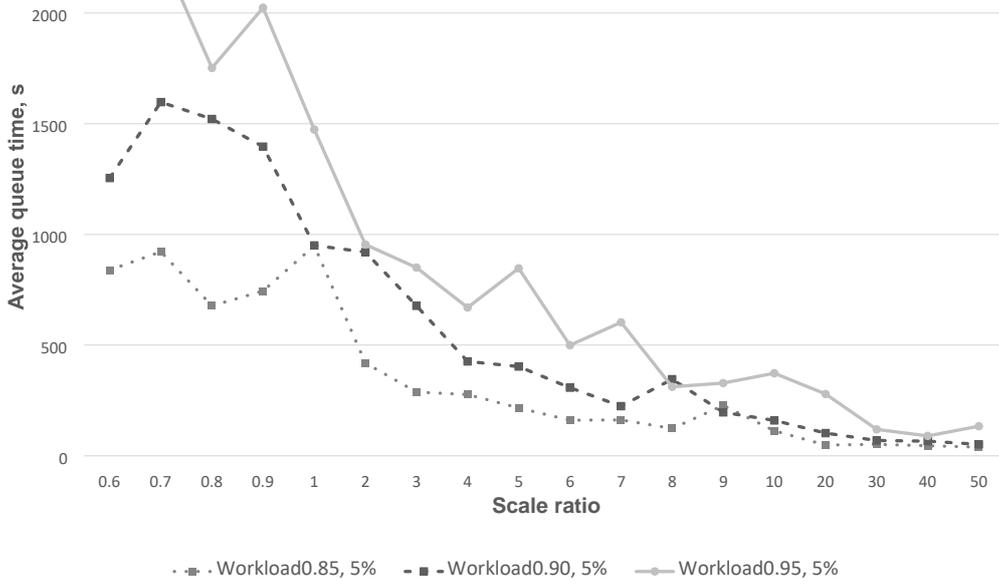

Figure 10. Average queue time for all Workload for 5% initialization proportion. The less the better

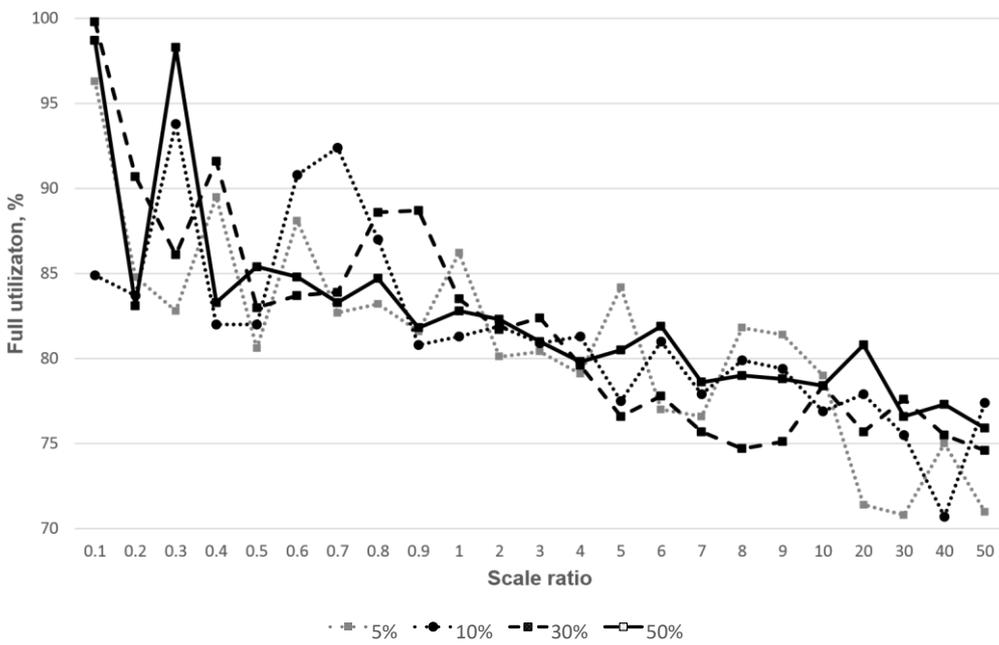

Figure 11. Full utilization for Workload0.85 for initialization proportion from 5% to 50%. The more the better

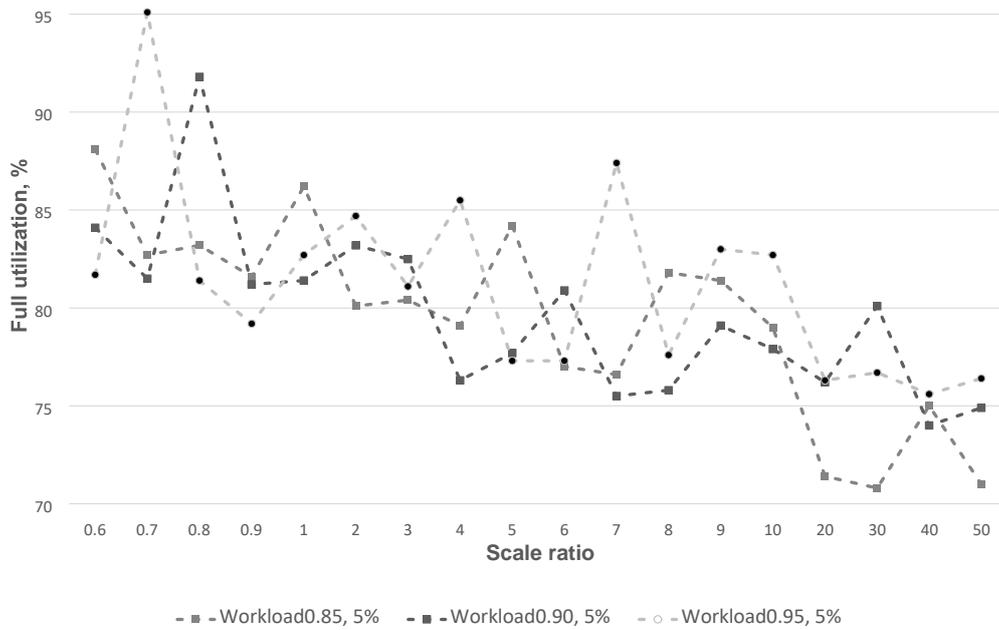

Figure 12. Full utilization for all Workload for 5% initialization proportion. The more the better

Let us fix the average initialization proportion. We study the influence of the workflow intensity on useful utilization. Figure 14 presents a comparative graph of useful utilization for 3 workflows with different intensities and 5% average initialization proportion. With an increase in the scale ratio no significant impact on the useful utilization of computing resources was detected.

As the scale ratio increases, the full utilization of computing resources varies within a narrow range of values, which is comparable to the fluctuation of this indicator during the operation of the JMS. There was no statistically significant effect of the scale ratio useful utilization detected.

## 8. CONCLUSIONS

The Alea-based model with implemented Packet algorithm significantly increases the range of explored scale ratio values without losing accuracy. In result, it was possible to evaluate the influence of the variety of parameters on efficiency metrics, including the influence of the scale ratio. Therefore, if we change the workflow, it is possible to perform a new study in an acceptable time and determine the best JMS settings.

In this study we conducted over a thousand experiments to examine different values of the scale ratio (from 0.1 to 1 with step of 0.1, from 1 to 10 with step of 1, from 10 to 100 with step of 10, and from 100 to 1000 with step of of 100), and different values of the average initialization proportion (from 5% to 50%), for different workflows. We used workflows that were generated using Lublin and Feitelson generator as well as workflows created by a modified generator that provided more homogeneous ones. For each of the generators we used 3 workloads with different calculated load: 0.85 for underload mode, 0.9 for loaded mode, and 0.95 for overload mode. At the same time, we found no significant impact of the homogeneity or intensity of the workflow on the scale ratio influence on the efficiency metrics. Intensity and heterogeneity only influence the absolute values of efficiency metrics.

The number of jobs in the queue, the average and median queue time decrease as the scale ratio increases to a certain value and stabilizes at it. Further increase of the scale ratio has no effect.

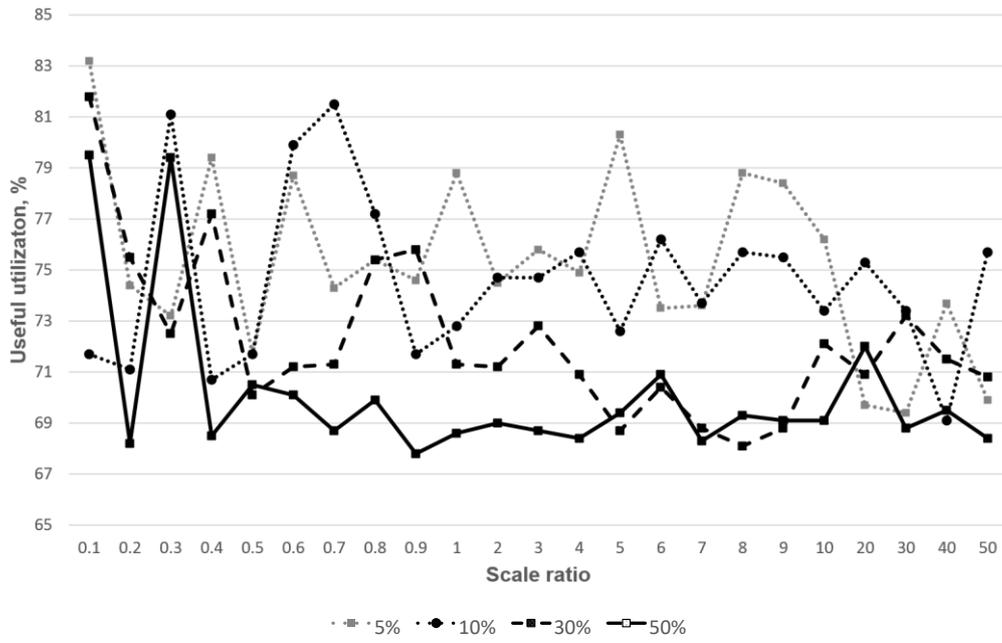

Figure 13. Useful utilization for Workload0.85 for different initialization proportions from 5 to 50%. The more the better

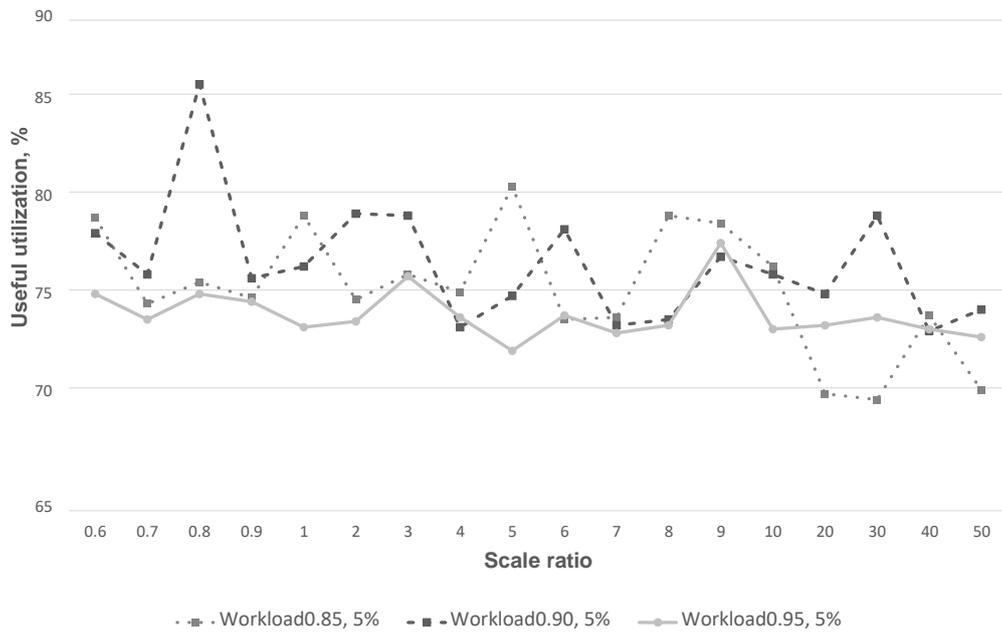

Figure 14. Useful utilization for all Workload for 5% initialization proportion. The more the better

The greatest value of full utilization is achieved at low values of the scale ratio. As the scale ratio increases, full utilization decreases. The influence of the scale ratio on useful utilization has not been identified; fluctuations of natural changes in utilization were observed.

For the known workload characteristics, the workload generator allow to obtain workflows for research. After doing that, the use of the Alea-based model makes it possible to determine the threshold value of the scale ratio when plateau of the average and median queue time is reached. Changing the scale ratio above the found threshold has no sense, since it does not influence the efficiency metrics. If a real workflow is available over a long period of time, a similar simulation can be carried out. It can be used instead of the generated workflow.

When tuning a grouping system, we recommend the JMS administrator to choose a balance between the required efficiency metrics.

When workflow properties are similar to investigated ones, then the administrator can use our data and set the appropriate scale ratio. The research results show that influence of scale ratio values on efficiency metrics does not essentially depend on intensity or homogeneity of workflow.

When the workflow properties differ significantly it is necessary to make new simulation using the presented JMS model. The simulation results will allow the administrator to find the scale ratio with acceptable balance between various efficiency metrics.